\documentclass[11pt,5p,twocolumn,preprint]{elsarticle}

\journal{~}

\usepackage[utf8]{inputenc}		%
\usepackage[T1]{fontenc}		%
\usepackage{graphicx}			%
\usepackage{subcaption}
\usepackage{amssymb} 			
\usepackage{amsmath} 			
\usepackage{wasysym} 

\usepackage{lineno}

\begin{document}

\begin{frontmatter}

\title{A phase-space description of the Earth System in the Anthropocene}

\author{O. Bertolami}
\ead{orfeu.bertolami@fc.up.pt}

\author{F. Francisco}
\ead{frederico.francisco@fc.up.pt}

\address{Departamento de Física e Astronomia and Centro de Física do Porto, Faculdade de Ciências, Universidade do Porto, Rua do Campo Alegre 687, 4169-007 Porto, Portugal}

\begin{abstract}
	Based on a dynamic systems approach to the Landau-Ginzburg model, a phase space description of the Earth System (ES) in the transition to the Anthropocene is presented. It is shown that, for a finite amount of human-driven change, there is a stable equilibrium state that is an attractor of trajectories in the system's phase space and corresponds to a Hothouse Earth scenario. Using the interaction between the components of the ES, it is argued that, through the action of the Technosphere, mitigation strategies might arise for which the deviation of the ES temperature from the Holocene average temperature is smaller.
\end{abstract}


\end{frontmatter}


\section{Introduction}

Recently, we have proposed that the Earth System (ES) transition from the Holocene to other stable geological eras was similar to a phase transition and could be described by the Landau-Ginzburg theory (LGT) \cite{Bertolami:2018}. This description suggests that the relevant thermodynamic variable to specify the state of the ES is the free energy, $F$, and that a relevant order parameter, $\psi$, is the relative temperature deviation from the Holocene average temperature, $\langle T_{\rm H}\rangle$, $\psi = (T - \langle T_{\rm H}\rangle)/ \langle T_{\rm H}\rangle$.

This physical framework allows for establishing the state of the ES in terms of the relevant physical variables, $(\eta, H)$, where $\eta$ corresponds to the astronomical, geophysical and internal dynamical causes of change, while $H$ stands for the human activities.

The Landau-Ginzburg model allows for obtaining the so-called Anthropocene equation, \textit{i.e.}, the evolution equation of the ES once it is dominated by the human activities, and also to show that the transition from the Holocene conditions to the Anthropocene are indeed associated to a great acceleration of the human activities. This is consistent with the empirical observation that the Anthropocene started by the second half of the 20th century \cite{Steffen:2014}.

In the present work we perform a phase space analysis of the temperature field, $\psi$, and show that the recently discussed Hothouse Earth scenario \cite{Steffen:2018} corresponds, for a finite amount of human driven change, to a stable minimum and, therefore, an attractor of the trajectories of the dynamical system that describes the ES in terms of the Landau-Ginzburg model. We also discuss how to ``engineer'' other possible minima for the ES.

This paper is organized as follows: in the next section we review the Landau-Ginzburg model proposal, discuss the Anthropocene Equation (AE), the dynamical system emerging from this description and its phase space. In section \ref{sec:hothouse}, the Hothouse earth state is shown to be an attractor of trajectories in the phase space of the ES and we discuss how to engineer new minima using new terms in the free energy. We also discuss the interactions between different terms of the planetary boundaries \cite{Steffen:2015}. This reinforces the importance of maintaining the ES within the so called \emph{safe operating space} \cite{Rockstrom:2009,Steffen:2011}. Finally, in section 4 we present our conclusions.


\section{The Anthropocene Equation phase-space}

In general terms, a dynamical system is any system that evolves in time. It is mathematically described by an ordinary differential equation (ODE) of the form $\dot{x} = f(x,t)$, where $\dot{x} = {dx/dt}$, $t$ is the time and $x$ represents a state of the system. The phase space $(x,\dot{x})$, for a dynamical system, is fully specified by the space of coordinates, $x(t)$, which hence contains all possible states of the system. An important class of dynamical systems, called autonomous, do not explicitly depend on time and their evolution equation has the form $\dot{x} = f(x)$. A dynamical system can also depend on parameters, $\alpha$, with an evolution equation of the form $\dot{x} = f(x,\alpha)$.

For a given set of initial conditions, corresponding to a state $x_0$ in the phase space, we can establish and solve the initial value problem with the evolution equation. Its solution is a function $x(t)$ describing the trajectory or \emph{orbit} of the dynamical system in the phase space.

The study of dynamical systems reveals many of its properties, notably the existence of attractors, without the need for solving explicitly all relevant initial value problems.

Dynamical systems in physics are described either through a Lagrangian function and the least action principle or the equivalent Hamiltonian formulation, whether it is in classical, statistical or quantum problems.

The dynamics of the system is encapsulated in a Lagrangian function, $\mathcal{L}(q,\dot{q},t)$, of a set of generalized coordinates of the system, $q$, their time derivatives, $\dot{q}$ and time itself, $t$. The time integral of the Lagrangian is a quantity called action and the minimization of the action amounts to determining the evolution equations of the system, the so-called Euler-Lagrange equations. Equivalently, we can derive a Hamiltonian function, $\mathcal{H}(q,p,t)$, where $p$ is the canonical conjugate momentum, which is defined below. In conservative physical applications, $\mathcal{H}(q,p,t)$ represents the total energy of the system, and lead to evolution equations equivalent to the Euler-Lagrange equations.

In Ref.\,\cite{Bertolami:2018}, we have introduced the free energy of the ES near a phase transition as described by the LGT:
\begin{equation}
	F(\eta,H) = F_0 + a(\eta)\psi^2 + b(\eta)\psi^4 - h(\eta)H\psi.
	\label{eq:free_energy}
\end{equation}
Despite its phenomenological nature, this description of the free energy, which is a thermodynamic potential, fits quite well in the Hamiltonian formalism to describe the dynamics of the ES.

The Hamiltonian system is given in terms of a set of canonical coordinates and conjugate canonical momenta. The Hamiltonian function is defined as
\begin{equation}
	\mathcal{H}(q, p) = p \dot{q} - \mathcal{L}(q,\dot{q}),
\end{equation}
where $p = {\partial \mathcal{L} / \partial \dot{q}}$ is the canonical conjugate momentum to the canonical coordinate $q$ and $\mathcal{L}(q.\dot{q})$ is the ES Lagrangian. We have already omitted the time dependence of the Lagrangian since we have assumed that the free energy does not depend explicitly on time.

The most general set of canonical coordinates for the ES in this description should include, not only the order parameter $\psi$, but also the natural and human drivers, $\eta$ and $H$, respectively, thus $q=(\psi,\eta,H)$.

In addition to the potential which we have already identified with the free energy, the Lagrangian should include a set of kinetic terms for the canonical coordinates. The simplest possible kinetic term is a quadratic term proportional to the squared first derivative of each coordinate. This way, we write the Lagrangian as
{\setlength\arraycolsep{2pt}
\begin{eqnarray}
	\mathcal{L}(q,\dot{q}) &=& \frac{\mu}{2} \dot{\psi}^2 + \frac{\nu}{2} \dot{\eta}^2 - F_0 \nonumber \\
	& & - a(\eta)\psi^2 - b(\eta)\psi^4 + h(\eta)H\psi,
\end{eqnarray}}
where $\mu$ and $\nu$ are constants.

We have argued that during the Anthropocene the ES is dominated by the effects of human activities. In terms of a dynamical systems description, this means that these have a much larger and faster effect than the longer time scales of natural factors. For this reason, for a study of the ES centred in the last century and currently we can safely drop the term in $\dot{\eta}^2$. However, any analysis of the ES for geological time-scales would have to include this term as well.

Notice that we could have introduced a kinetic term for the human activities, $\dot{H}^2$. This term could describe how the ES at large affects the human activities themselves. Although this feedback loop exists, we shall assume instead that $H$ is an external force.

We shall start with a simplified analysis focusing only on the order parameter used in the LGT formulation, $\psi$, representing temperature, undoubtedly the key thermodynamic and climatic state variable. The additional terms may become relevant as more encompassing descriptions of the ES arise. Our Lagrangian, simplified along this reasoning, becomes  
\begin{equation}
	\mathcal{L}(\psi,\dot{\psi}) = \frac{\mu}{2} \dot{\psi}^2 - a \psi^2 - b \psi^4 + h H\psi,
\end{equation}
where $a$, $b$ and $h$ are constants (\textit{cf.} Ref.\,\cite{Bertolami:2018}) and we have dropped the constant $F_0$ since it will not affect the dynamics of the system.

The canonical momentum can be then obtained as
\begin{equation}
	p = \frac{\partial \mathcal{L}}{\partial \dot{\psi}} = \mu \dot{\psi}.
\end{equation}
We now have all the ingredients to write the Hamiltonian of the ES
\begin{equation}
	\mathcal{H}(\psi, p) = \frac{p^2}{2\mu} + a \psi^2 + b \psi^4 - h H\psi.
\end{equation}

Finally, we can use Hamilton's equations
\begin{equation}
	\dot{\psi} = \frac{\partial \mathcal{H}}{\partial p}, \quad \dot{p} = - \frac{\partial \mathcal{H}}{\partial \psi}.
\end{equation}
to obtain the evolution equation of the dynamical system in the phase space $(\psi,\dot{\psi})$:
\begin{equation}
	\dot{\psi} = \frac{p}{\mu}, \quad \dot{p} = - 2 a \psi - 4 b \psi^3 + h H.
	\label{eq:evolution_eqs}
\end{equation}

\begin{figure}
	\centering
	\includegraphics[width=\columnwidth]{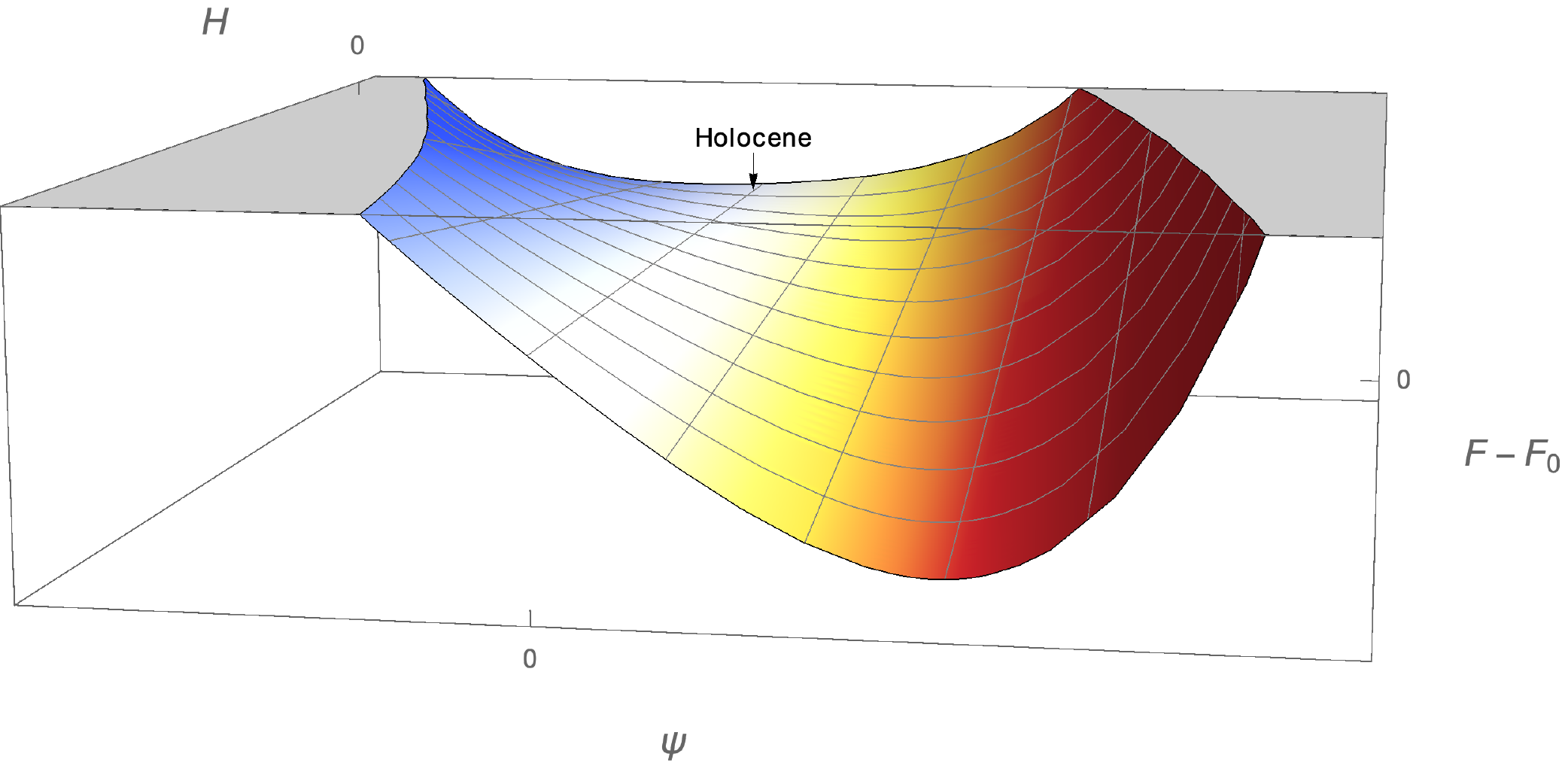}
	\caption{Stability landscape of the ES in terms of $\psi$ and $H$.}
	\label{fig:stability_landsacpe}
\end{figure}

With these equations, we can plot the phase space portrait of the dynamical system, allowing us to graphically identify the orbits and the attractors. To start with, we should have in mind the stability landscape of the temperature field, $\psi$, depicted in Figure\,\ref{fig:stability_landsacpe}. In the center of the valley for $H=0$ we have the Holocene minimum described in Ref.\,\cite{Bertolami:2018}. Moving away from $H=0$, we clearly see that the human intervention opens up a deeper, and therefore more stable, hotter minimum that we can identify with the Hothouse Earth \cite{Steffen:2018}.

\begin{figure}
	\centering
    \includegraphics[width=0.85\columnwidth]{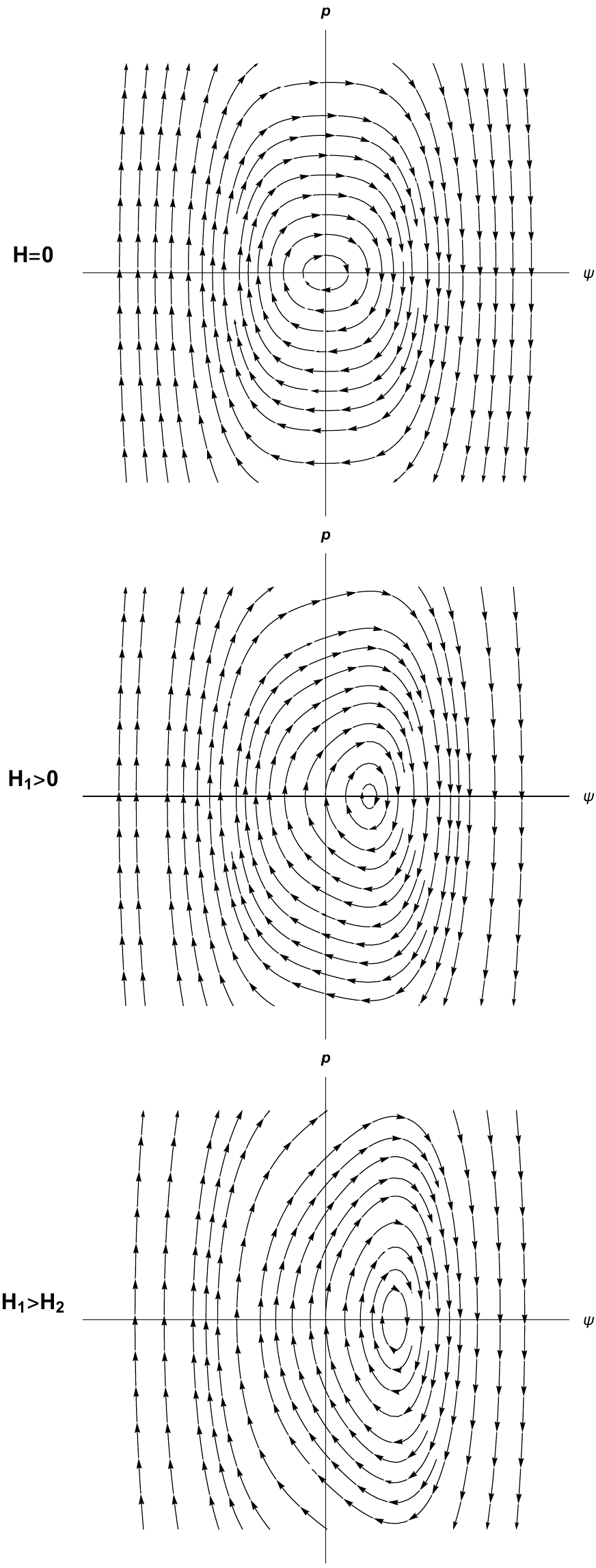}
	\caption{Phase portrait examples of the ES dynamical system as obtained from Eq.\,(\ref{eq:evolution_eqs})}
	\label{fig:phaseportrait}
\end{figure}

We can now examine the possible trajectories given the dynamical system, Eq.\,(\ref{eq:evolution_eqs}). As mentioned, in this description of the system, $\eta$ is fixed and $H$ is treated as a parameter with a temporal dependence. The ES dynamics is reflected in the phase space and, thus, on the position and strength of its attractors, as exemplified in Figure\,\ref{fig:phaseportrait} for increasing values of $H$.

We can obtain the dynamical system orbits analytically for a simpler case, but which is still useful to to acquire some understanding about the phase space.

If $b \simeq 0$, we can drop the cubic term in Eq.\,(\ref{eq:evolution_eqs}), the equations of motion now correspond to a harmonic oscillator with an external force $H(t)$.

We further consider a specific function of time for the external force, in this case $H(t) = H_0 t$. Our simplified system now has equations of motion of the form
\begin{equation}
	\mu \ddot{\psi} = - 2 a \psi + h H_0 t.
\end{equation}
If we assume a departure from equilibrium, then $\dot{\psi}(0)=0$ and the solution is given by
\begin{equation}
	\psi(t) = \psi_0 \cos (\omega t) + \alpha t,
\end{equation}
where $\omega = \sqrt{2a / \mu} $ is an angular frequency, $\alpha = h H_0 / 2 a$ and $\psi_0$ is an arbitrary constant fixed by the initial conditions of the problem.

It can be shown that this solution corresponds to elliptical trajectories in the phase space with moving foci of the form
\begin{equation}
	{{\Psi}^2 \over \psi_0^2 } + {\dot{\Psi}^2 \over \psi_0^2 \omega^2} = 1,
\end{equation}
where $\Psi = \psi + \alpha t$. 

\begin{figure}
	\centering
	\includegraphics[width=0.8\columnwidth]{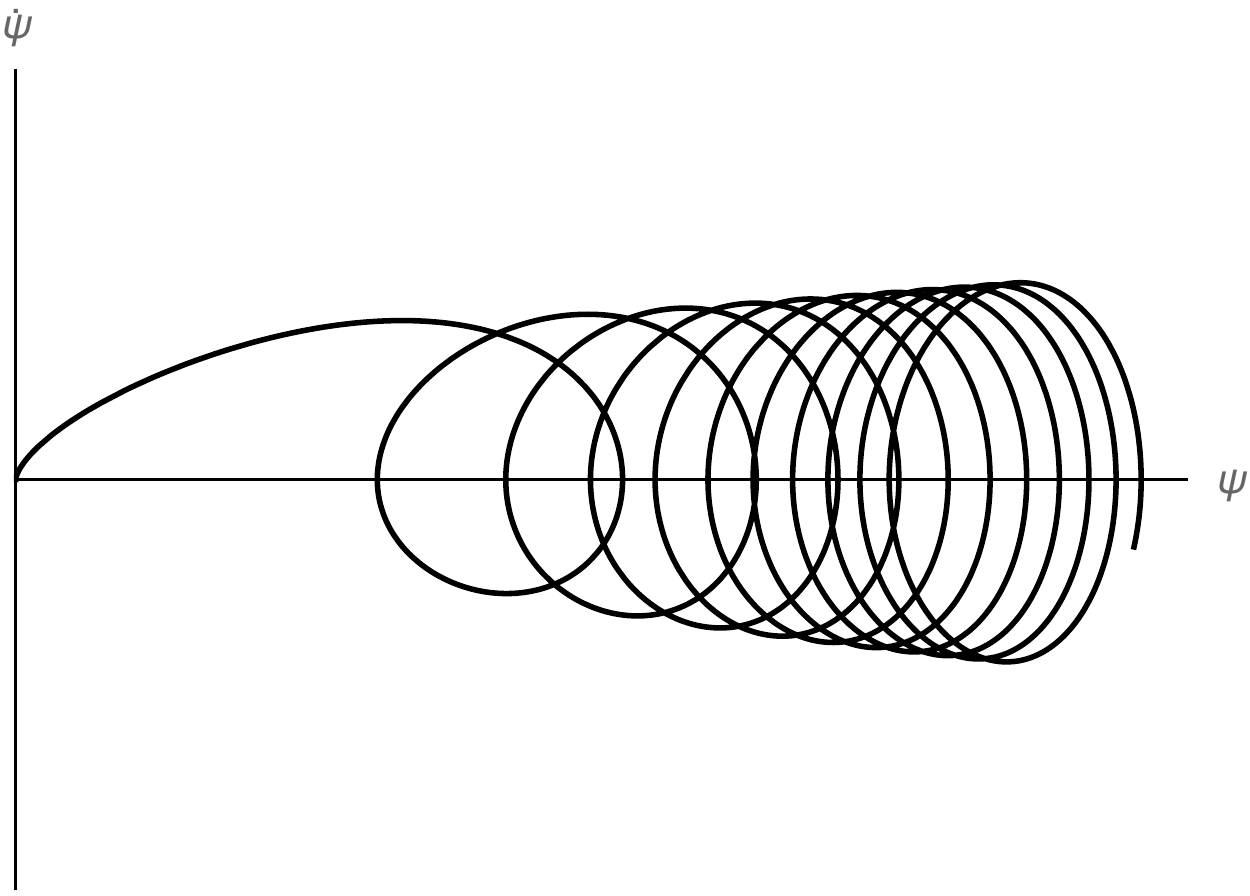}
	\caption{Trajectory of the ES in the phase space for $H(t)=t$ and initial conditions $(\psi,p)(0)=(0,0)$, modelling the departure from the Holocene equilibrium.}
	\label{fig:EStrajectory_phasesp}
\end{figure}

These results provide a qualitative picture that still holds after the cubic term is reintroduced. Its effect is to slightly deform the ellipse and to slow its movement to towards higher values of $\psi$. When we numerically solve the equations of motion, Eq.\,(\ref{eq:evolution_eqs}), we can then plot the trajectory of the dynamical system for any given initial conditions. An example is shown in Figure\,\ref{fig:EStrajectory_phasesp}. Notice that our analysis could be generalized for the case where $H(t)$ has a quadratic, cubic or, indeed, any time dependence.

\begin{figure}
	\centering
	\includegraphics[width=0.7\columnwidth]{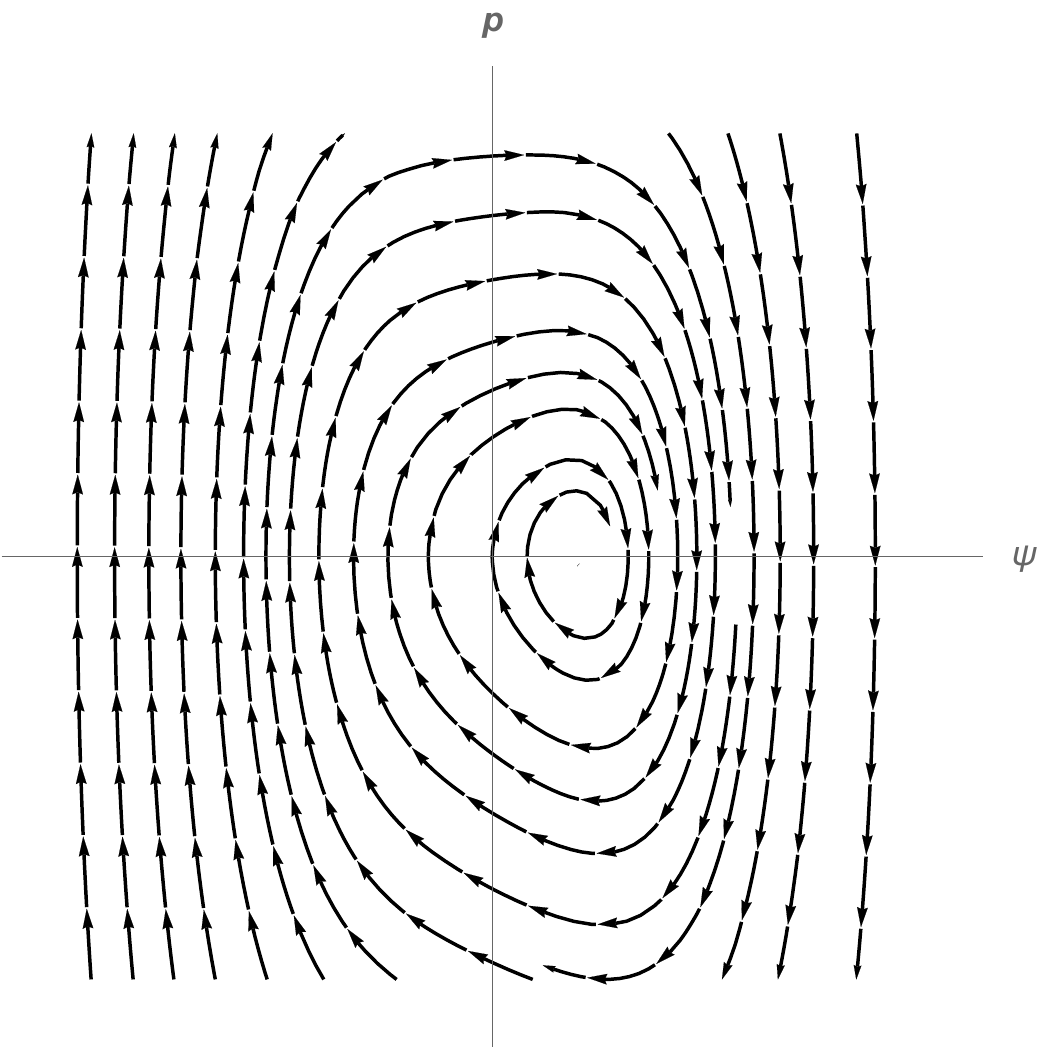}
	\caption{Phase portrait example of the Earth dynamical system with a Rayleigh dissipative term as obtained from Eq.\,(\ref{eq:evolution_eq_dissipative}). Compare with the central panel in Figure\,\ref{fig:phaseportrait}.}
	\label{fig:phaseportrait_damp}
\end{figure}

Still, our model so far presents the picture of an ES perpetually orbiting around the equilibrium point. This is due to the fact that our description up to this point is fully conservative.

All real physical systems have some dissipative character that ensures the system always moves towards and eventually reaches a stable equilibrium. In thermodynamics, this is equivalent to an increase in the entropy, $S$, which we know should always grow with time. Our description of the ES is based on its free energy, $F = U - TS$, where $U$ is internal energy and $T$ is the thermodynamic temperature. If the internal energy remains constant, then the second law of thermodynamics ensures that the system will move towards its minimum attainable free energy.

The simplest way to insert a dissipative term in the Lagrangian formalism is through a Rayleigh function, $R = {1 \over 2} k \dot{\psi}^2$, where $k$ is a positive constant, which gives a contribution to the equation of motion $\partial R / \partial \dot{\psi}$. With this additional term, Eq.\,(\ref{eq:evolution_eqs}) becomes
\begin{equation}
	\mu \ddot{\psi} = - 2 a \psi - 4 b \psi^3 + h H - k \dot{\psi},
	\label{eq:evolution_eq_dissipative_2}
\end{equation}
or, in terms of the momentum $p$,
\begin{equation}
	\dot{p} = - 2 a \psi - 4 b \psi^3 + h H - {k \over \mu} p.
	\label{eq:evolution_eq_dissipative}
\end{equation}

Notice that Eq.\,(\ref{eq:evolution_eq_dissipative_2}) (or Eq.\,(\ref{eq:evolution_eq_dissipative})) could itself be regard as an Anthropocene equation, as its solutions allow for predicting  the temperature field for a given evolution of the human forcing. However, in the literature, the designation Anthropocene equation is reserved for the evolution equation of the ES itself, which was obtained in Ref.\,\cite{Bertolami:2018} through the LGT.

The phase portrait of the dynamical system with a dissipative term shows the orbits changing from deformed ellipses to spirals towards the free energy minimum, as depicted in Figure\,\ref{fig:phaseportrait_damp}.

The existence of a stable equilibrium points in dynamical systems can be shown through the Lyapunov theorem.

A critical point $(\psi_{\rm c}, p_{\rm c})$ is said to be \emph{Lyapunov stable} if any trajectory starting with a given neighbourhood of that point remains within a finite neighbourhood of it. The orbits of the ES in Figure\,\ref{fig:phaseportrait} for each constant value of $H$ are Lyapunov stable around the minimum of the free energy.

If, additionally, there exists a finite neighbourhood of the critical point within which all trajectories converge to the critical point, then that critical point is said to be \emph{asymptotically stable}. When a dissipative term is added to the ES evolution equation with constant $H$, the critical points in the minimum of the free energy becomes asymptotically stable. In this case, the critical point is also an \emph{attractor} of trajectories.

\begin{figure}
	\centering
	\includegraphics[width=0.8\columnwidth]{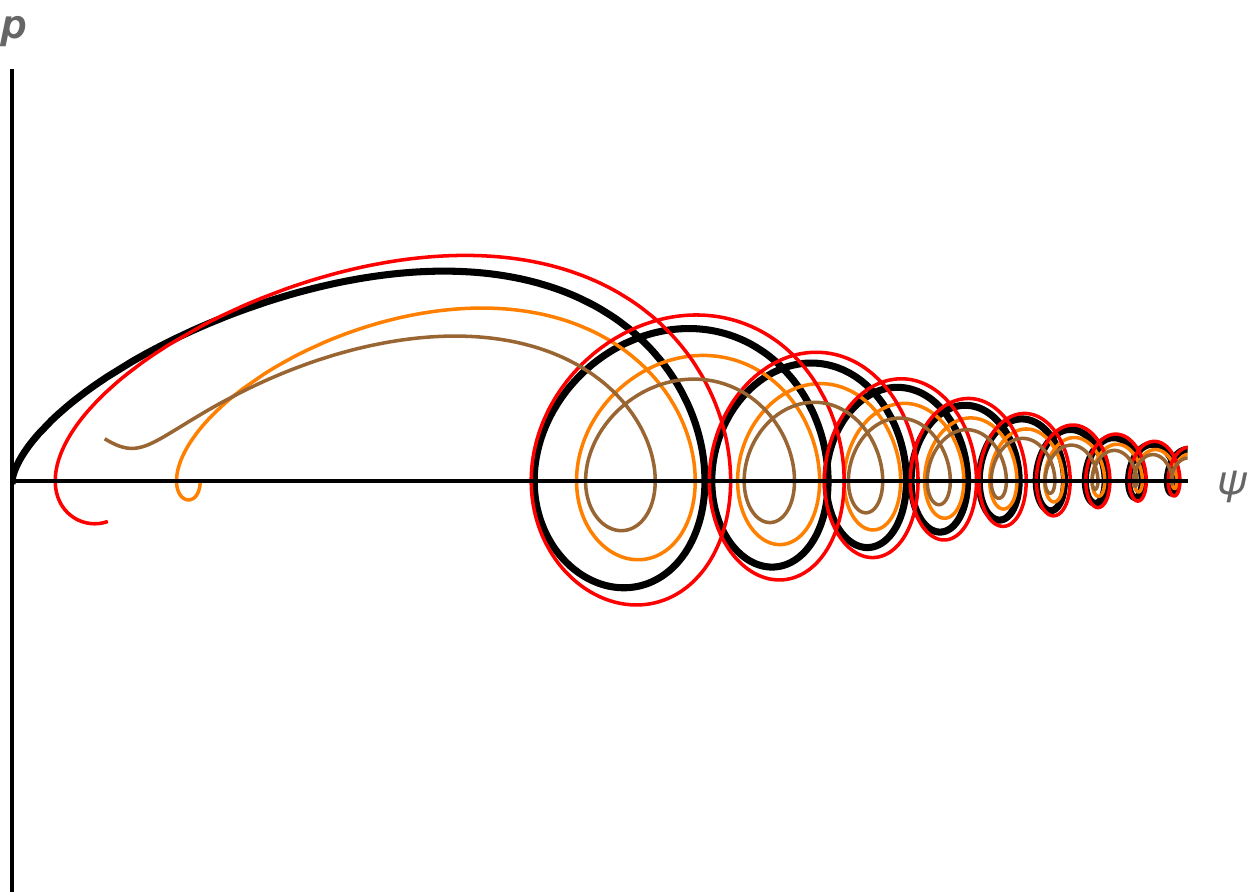}
	\caption{Trajectory of the ES in the phase space for $H(t)=t$ and initial conditions $(\psi,p)(0)=(0,0)$ (black line), modelling the departure from the Holocene equilibrium, with the inclusion of a Rayleigh dissipative term in the Lagrangian. Gray lines show evolution for different initial conditions to show how they all converge to the same moving equllibrium point.}
	\label{fig:EStrajectory_phasesp_damp}
\end{figure}

In the scenario where $H$ depends on time, this will cause a continuous shift of the critical point that causes the kind of trajectories exemplified in Figures\,\ref{fig:EStrajectory_phasesp} and \ref{fig:EStrajectory_phasesp_damp}, respectively, with and without a dissipative term. Strictly speaking, the system is only stable if $H(t)$ is bounded. The Lyapunov condition is ensured for our system as, from Eq.\,(\ref{eq:evolution_eq_dissipative}), we can see that $p(0) \leq hH - (k/\mu) p(0)$ for $\psi > 0$, implying that there is a state $(\psi_{\rm c}, p_{\rm c})$ for which $p_{\rm c} = 0$, for a finite $H$. In that case, the ES will converge to that new critical point with $\psi_{\rm c} >0$, corresponding to the Hothouse Earth state described in Ref.\,\cite{Steffen:2018}.

\begin{figure}
	\centering
	\includegraphics[width=0.8\columnwidth]{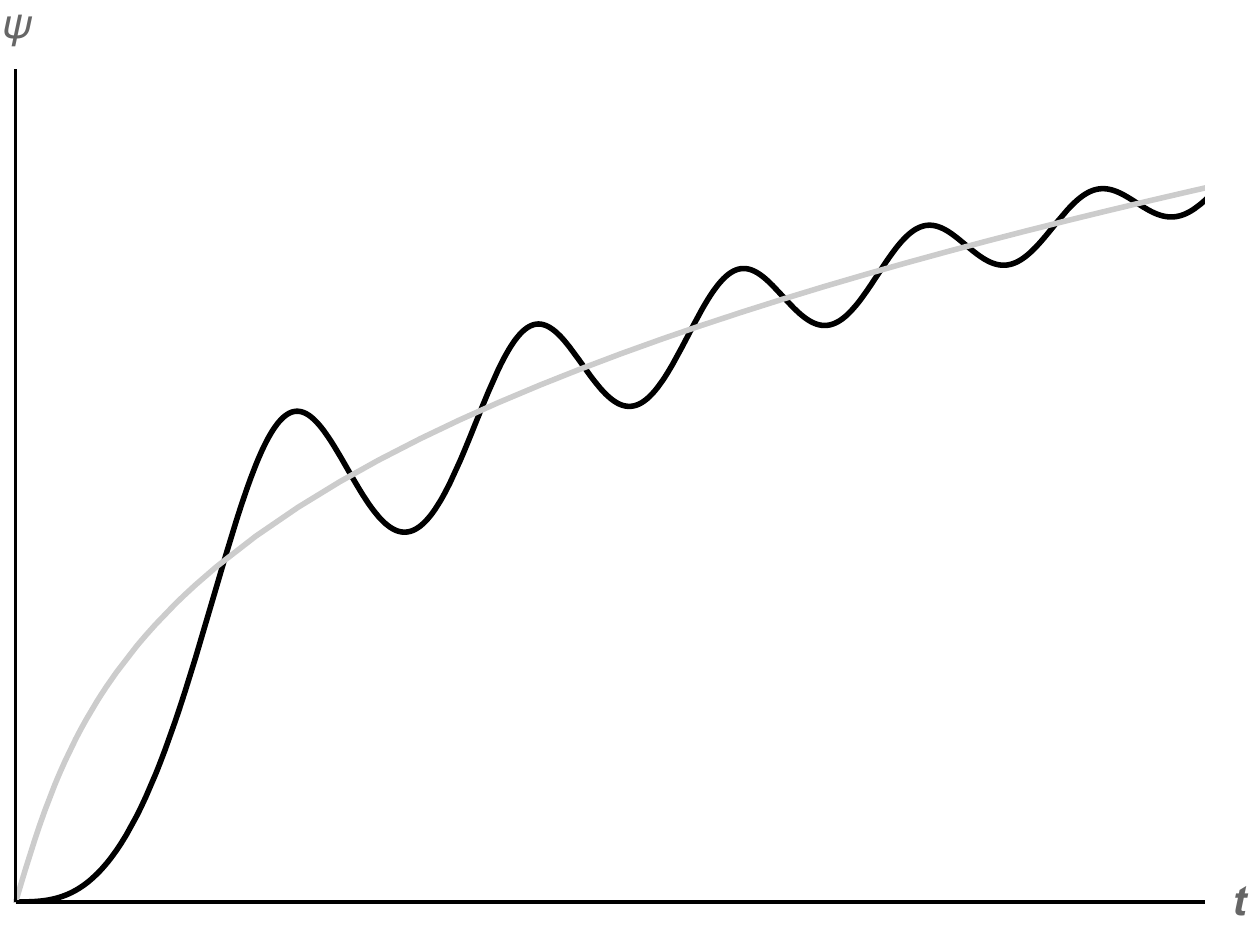}
	\caption{Evolution of the ES temperature, $\psi$, and a function of time for $H(t)=t$ and initial conditions $(\psi,p)(0)=(0,0)$ (black line), modelling the departure from the Holocene equilibrium, with the inclusion of a Rayleigh dissipative term in the Lagrangian. The gray line represents the evolution of the equilibrium point.}
	\label{fig:EStemperature_time}
\end{figure}

The actual evolution of the temperature (black oscillating line) is shown in Figure\,\ref{fig:EStemperature_time}. The gray curve corresponds to the associated equilibrium state which evolves as $\langle \psi \rangle \sim H^{1 \over 3}$ \cite{Bertolami:2018}. Hence, if the effect of the human activities led to an increase of $1\,{\rm K}$ since the beginning of the Anthropocene \cite{Jones:2004}, about 50 years ago, then 100 years after the start of the Anthropocene we can expect a temperature increase to $\sqrt[3]{2} = 1.26\,{\rm K}$. This is, of course, assuming that the growth of $H$ remains linear.

\section{Hothouse Earth state and other minima}
\label{sec:hothouse}

We have seen in the previous section that a critical point of the dynamical system corresponds necessarily to an ES trajectory towards a minimum where the temperature is greater that the one at the Holocene equilibrium.

As discussed in Ref.\,\cite{Steffen:2018}, this increase in the global temperature can lead to a chain failure of the main regulatory ecosystems of the ES that already show tipping point features. It is therefore quite relevant to investigate the possibility of engendering alternative trajectories to the ES. This can be achieved through the engineering of metastable minima on the phase space, that is, new terms proportional to $\psi^2$ and/or $\psi^3$ in the free energy function, Eq.\,(\ref{eq:free_energy}) that force the ES to remain close to the Holocene minimum.

Another strategy would be to consider the change in the sign of the human drivers, $H$. In the context of the planetary boundaries framework \cite{Steffen:2015}, the state of the ES is specified through a set of 9 parameters and the Holocene-like conditions are ensured provided the ES remains within the so-called Safe Operating Space (SOS) \cite{Rockstrom:2009}.

Indicating the effect of the human activity in altering the optimum Holocene conditions by $h_i$, the bulk of the human intervention, $H$, can be written as
\begin{equation}
	H = \sum_{i=1}^{9} h_i + \sum_{i,j=1}^{9} g_{ij} h_i h_j + \sum_{i,j,k=1}^{9} \alpha_{ijk} h_i h_j h_k + \ldots,
\end{equation}
where the second and third set of terms indicate the interaction between the various effects of the human action on the planetary boundary parameters. Of course, higher order interactions terms can be considered, but we shall restrict our considerations up to second order and, in fact, to a subset of planetary boundary parameters. It is physically reasonable and mathematically convenient to assume that the $9 \times 9$ matrix, $[g_{ij}]$ is symmetric, $g_{ij} = g_{ji}$, and non-degenerate, $\det[g_{ij}] \neq 0$.

Let us consider only a couple of parameters, say $h_1$ and $h_9$, and assume, in particular, that the ninth parameter corresponds to the Technosphere, \textit{i.e.}, the set of human technological activities aiming to repair or to mitigate the action on the variables away from the SOS. Thus, under these conditions, we can write $H$ as
\begin{equation}
	H = h_1 + h_9 + 2 g_{19} h_1 h_9 + g_{11} h_1^2 + g_{99} h_9^2.
\end{equation}

It is easy to see that if $h_1 > 0$, $h_9 < 0$ and $g_{19} > 0$, then the effect is to mitigate the destabilizing effect of $h_1$. The net effect of this technological interaction is to ensure that the minimum due to human intervention is, as discussed in Ref.\,\cite{Bertolami:2018}, closer to the Holocene, minimising $\langle \psi \rangle$, given that $\langle \psi \rangle$ is proportional to the cubic root of $H$. We could also argue that $g_{11}$ and $g_{99}$ are negative too, given they can have an inhibiting effect on affecting themselves.

\begin{figure}
	\centering
	\includegraphics[width=\columnwidth]{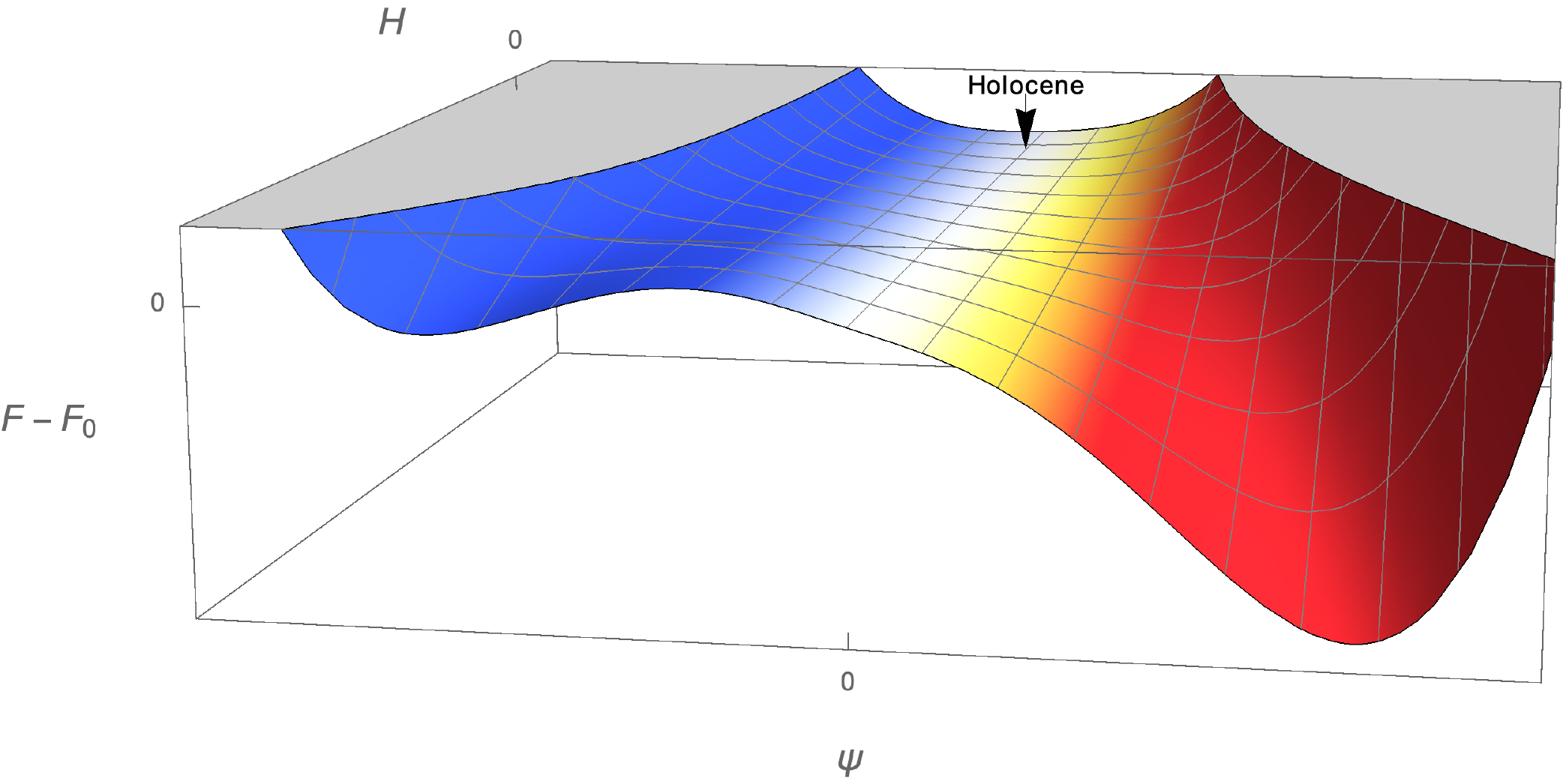}
	\caption{Stability landscape of the ES in terms of $\psi$ and $H$ with a metastable state.}
	\label{fig:stability_landsacpe_metastable}
\end{figure}

The above considerations assume that the $h_i$ terms do not depend on the temperature field. However, it is most likely, in physical terms, that $h_i = h_i(\psi)$, meaning in fact that the effect of the human activities might alter the free energy introducing new quadratic and cubic terms. The effect of these terms would depend, of course, on their relative strength, but may lead to the appearance of a metastable state, as illustrated on the left side of the plot in Figure\,\ref{fig:stability_landsacpe_metastable} for a $\psi^3$ type term.

\section{Conclusions}

Starting from the physical framework provided by the Landau-Ginzburg theory, we have built a description of the ES transition from the Holocene to the Anthropocene as a dynamical system. That way, we have a mathematical model well grounded on physics that can capture the behaviour of the ES during this transition, including its non-linearities and more complex properties.

We thus use the Hamiltonian formulation, ubiquitous in most branches of physics, to obtain the evolution equations of the ES and examine its orbits in the phase space, the space of all possible states of the system. It then becomes evident how the increase in human activities progressively deforms the phase space of the ES towards higher temperatures.

With this kind of mathematical description of the ES we can start to look for some of its properties, namely the existence and stability of equilibrium points. Indeed, we show how a given evolution of the human influence will affect these points. Even assuming that their effects are bounded, the ES will unquestionably progress towards an equilibrium away from the Holocene, supporting a Hothouse Earth scenario.

A key issue remains in how to correctly describe the human effects in their multiple components. We stress how these components are not independent and may have direct (second-order) or higher order interactions. We show, at least theoretically, that interactions among these components, most particularly the one associated with the Technosphere, can allow for mitigating strategies in what concerns the inevitable evolution towards a Hothouse Earth, as well as the ``engineering'' of metastable states where the ES can remain temporarily in equilibrium in a cooler state.

\section*{Acknowledgements}

The authors would like to thank Will Steffen for the insightful discussions.


\section*{References}

\bibliographystyle{elsarticle-num}
\biboptions{sort&compress}

\bibliography{anthropocene_phase_space}

\end{document}